\documentclass[aps,prm,superscriptaddress,twocolumn,showpacs]{revtex4-1}
\usepackage{graphicx}
\usepackage{amsmath}
\usepackage{epsfig}
\usepackage{multirow}
\usepackage{array}
\usepackage{inputenc}
\usepackage{color}
\usepackage{tabularx}
\usepackage{textcomp}
\usepackage{ulem}
\usepackage[colorlinks=true,linkcolor=black, citecolor=blue, urlcolor=blue, 
    unicode=true]{hyperref}
\begin{document}


\title{Common acoustic phonon lifetimes in inorganic and hybrid lead halide perovskites}


\author{M. Songvilay}
\affiliation{School of Physics and Astronomy, University of Edinburgh, Edinburgh EH9 3FD, UK}
\author{N. Giles-Donovan}
\affiliation{Medical and Industrial Ultrasonics, School of Engineering, University of Glasgow G128QQ, UK}
\author{M. Bari}
\affiliation{Department of Chemistry and 4D LABS, Simon Fraser University, Burnaby, British Columbia, V5A1S6 Canada}
\author{Z.-G. Ye}
\affiliation{Department of Chemistry and 4D LABS, Simon Fraser University, Burnaby, British Columbia, V5A1S6 Canada}
\author{J. L. Minns}
\affiliation{School of Physical Sciences, Ingram Building, University of Kent, Canterbury, Kent CT2 7NH, UK}
\author{M. A. Green}
\affiliation{School of Physical Sciences, Ingram Building, University of Kent, Canterbury, Kent CT2 7NH, UK}
\author{Guangyong Xu}
\affiliation{NIST Center for Neutron Research, National Institute of Standards and Technology, 100 Bureau Drive, Gaithersburg, Maryland, 20899, USA}
\author{P. M. Gehring}
\affiliation{NIST Center for Neutron Research, National Institute of Standards and Technology, 100 Bureau Drive, Gaithersburg, Maryland, 20899, USA}
\author{K. Schmalzl}
\affiliation{Forschungszentrum J\"ulich GmbH, J\"ulich Centre for Neutron Science at ILL, 71 avenue des Martyrs, 38000 Grenoble, France}
\author{W. D. Ratcliff}
\affiliation{NIST Center for Neutron Research, National Institute of Standards and Technology, 100 Bureau Drive, Gaithersburg, Maryland, 20899, USA}
\author{C. M. Brown}
\affiliation{NIST Center for Neutron Research, National Institute of Standards and Technology, 100 Bureau Drive, Gaithersburg, Maryland, 20899, USA}
\author{D. Chernyshov}
\affiliation{Swiss-Norwegian Beam Lines, European Synchrotron Radiation Facility, Polygone Scientifique Louis N\'eel, 6 rue Jules Horowitz, 38000 Grenoble, France}
\author{W. van Beek}
\affiliation{Swiss-Norwegian Beam Lines, European Synchrotron Radiation Facility, Polygone Scientifique Louis N\'eel, 6 rue Jules Horowitz, 38000 Grenoble, France}
\author{S. Cochran}
\affiliation{Medical and Industrial Ultrasonics, School of Engineering, University of Glasgow G128QQ, UK}
\author{C. Stock}
\affiliation{School of Physics and Astronomy, University of Edinburgh, Edinburgh EH9 3FD, UK}
\date{\today}


\begin{abstract}

The acoustic phonons in the organic-inorganic lead halide perovskites have been reported to have anomalously short lifetimes over a large part of the Brillouin zone.  The resulting shortened mean free paths of the phonons have been implicated as the origin of the low thermal conductivity.   We apply neutron spectroscopy to show that the same acoustic phonon energy linewidth broadening (corresponding to shortened lifetimes) occurs in the fully inorganic CsPbBr$_{3}$ by comparing the results on the organic-inorganic CH$_{3}$NH$_{3}$PbCl$_{3}$ (Ref. \onlinecite{Songvilay2018}).  We investigate the critical dynamics near the three zone boundaries of the cubic $Pm\overline{3}m$ Brillouin zone of CsPbBr$_{3}$ and find energy and momentum broadened dynamics at momentum points where the Cs-site ($A$-site) motions contribute to the cross section.  Neutron diffraction is used to confirm that both the Cs and Br sites have unusually large thermal displacements with an anisotropy that mirrors the low temperature structural distortions.   The presence of an organic molecule is not necessary to disrupt the low-energy acoustic phonons at momentum transfers located away from the zone center in the lead halide perovskites and such damping may be driven by the large displacements or possibly disorder on the $A$ site.

\end{abstract}

\pacs{}

\maketitle


\section{Introduction}

The lead halide perovskites (with chemical formula $A$Pb(Cl,Br,I)$_{3}$) display exceptional structural, optical, electronic, and charge transport properties~\cite{Hirotsu1974,Babin1999,Stoumpos2013,Kondo2004,Ramade2018,Du2017}. In particular, the semiconducting properties of CsPbBr$_3$ have made this system a promising candidate for detector applications \cite{Stoumpos2013}. Recent theoretical and experimental works have been focused on hybrid organic-inorganic lead halide perovskites as potentially efficient photovoltaic materials \cite{Leguy2015,Egger2016,Saparov2016}. These materials combine two sublattices, formed by an inorganic lead halide framework coupled to a molecular cation (based on the $A$ site) through hydrogen bonding. As this coupling has been identified as key for understanding their improved efficiency in solar power conversion, a significant number of studies have been devoted to investigating the coupling between the molecular and inorganic frameworks \cite{Brown2017, Swainson2015, Ferreira2018, Songvilay2018} with some studies implicating the organic molecule as the origin for the improved thermal properties~\cite{GoldParker2018}. 

Although this new class of materials has generated a considerable amount of theoretical and experimental interest, it has been shown that all-inorganic lead halide perovskites also exhibit similar efficiencies and properties for photovoltaic devices \cite{Egger2018,Lee2017,Eperon2015,Ramasamy1959,Yantara2015}. Therefore, an open question is whether the organic A-site cation is essential to the enhanced photo-electronic properties~\cite{Kulbak2015}. From a more fundamental point of view, the role of the cationic nature and its influence on lattice dynamics is still unclear and needs to be further examined particularly given the suggestions that the low-energy acoustic phonons are influential to the electronic properties~\cite{Long2019}.  In this context, we investigate the low-energy acoustic phonons in CsPbBr$_{3}$ with the goal of establishing which lattice dynamical features are unique to the organic-inorganic variants. With Cs occupying the $A$-site, CsPbBr$_{3}$ is one of the all inorganic lead-halide perovskites.

The study of cesium lead halide compounds started 40 years ago when the first structural characterization was performed using neutron inelastic scattering in CsPbCl$_3$ by Fujii \textit{et al.}, who reported the temperature dependence of superlattice Bragg peaks through the structural transitions, as well as the dispersion of acoustic phonons across the Brillouin zone \cite{Fujii1974}. In an attempt to characterize the structural transitions in this compound, this study showed the presence of overdamped phonons at the $M$ and $R$ zone boundaries, which prevents the observation of any phonon softening near the transitions. A detailed study of the structural transitions was later performed by Hua \textit{et al.} \cite{Hua1991} who reported three structural transitions in CsPbCl$_3$: a cubic to tetragonal transition, followed by a tetragonal to orthorhombic phase transition and finally an orthorhombic-orthorhombic transition. 

A neutron diffraction study was also performed on the bromine compound \cite{Hirotsu1974} which exhibits only two structural transitions: one first order transition from a cubic phase described by the space group $Pm\overline{3}m$ to a tetragonal symmetry around 400~K (space group $P4/mbm$), and a second order transition to an orthorhombic phase around 360~K (space group $Pmbn$). The structural transitions in both chlorine and bromine compounds were then further investigated with ultrasound measurements that reported several anomalies in the longitudinal and transverse sound velocities at the structural transitions, indicative of changes in the elastic coefficients \cite{Hirotsu1977,Hirotsu1978}. Since then, many experimental and theoretical studies have been devoted to similar compounds, such as cesium tin halides \cite{Yang2017,Bechtel2018_1,Marronnier2017}.  

This paper reports a neutron scattering study of the transverse acoustic phonons in the all-inorganic perovskite CsPbBr$_3$ towards the three high symmetry points of the cubic $Pm\overline{3}m$  Brillouin zone $\textbf{Q}_X=(2,\frac{1}{2}, 0)$, $\textbf{Q}_M=(\frac{3}{2},\frac{1}{2}, 0)$, and $\textbf{Q}_R=(\frac{1}{2},\frac{1}{2}, \frac{3}{2})$. We show the existence of both critical fluctuations and resolution limited Bragg peaks at specific zone boundaries when approaching the orthorhombic transition, concurrently with  the presence of damped phonon modes indicating shortened lifetimes. Based on powder diffraction data showing strong anisotropic Cs and Br displacements in the tetragonal and orthorhombic phases, we speculate that the strong phonon damping originates from anharmonic effects related to this anisotropy. Based on these results, we discuss the importance of displacive transitions in this compound along with the influence of possible anisotropic $A$-site fluctuations affecting the acoustic phonon lifetime. This work therefore highlights similar mechanisms affecting the harmonic phonons as in the hybrid counterparts, which may explain their comparable efficiencies in terms of thermal and opto-electronic properties. This also illustrates that the strong phonon damping is possibly universal across the lead-halide perovskites and does not require the presence of an organic molecule on the $A$ site.

\begin{figure}
 \includegraphics[scale=0.31]{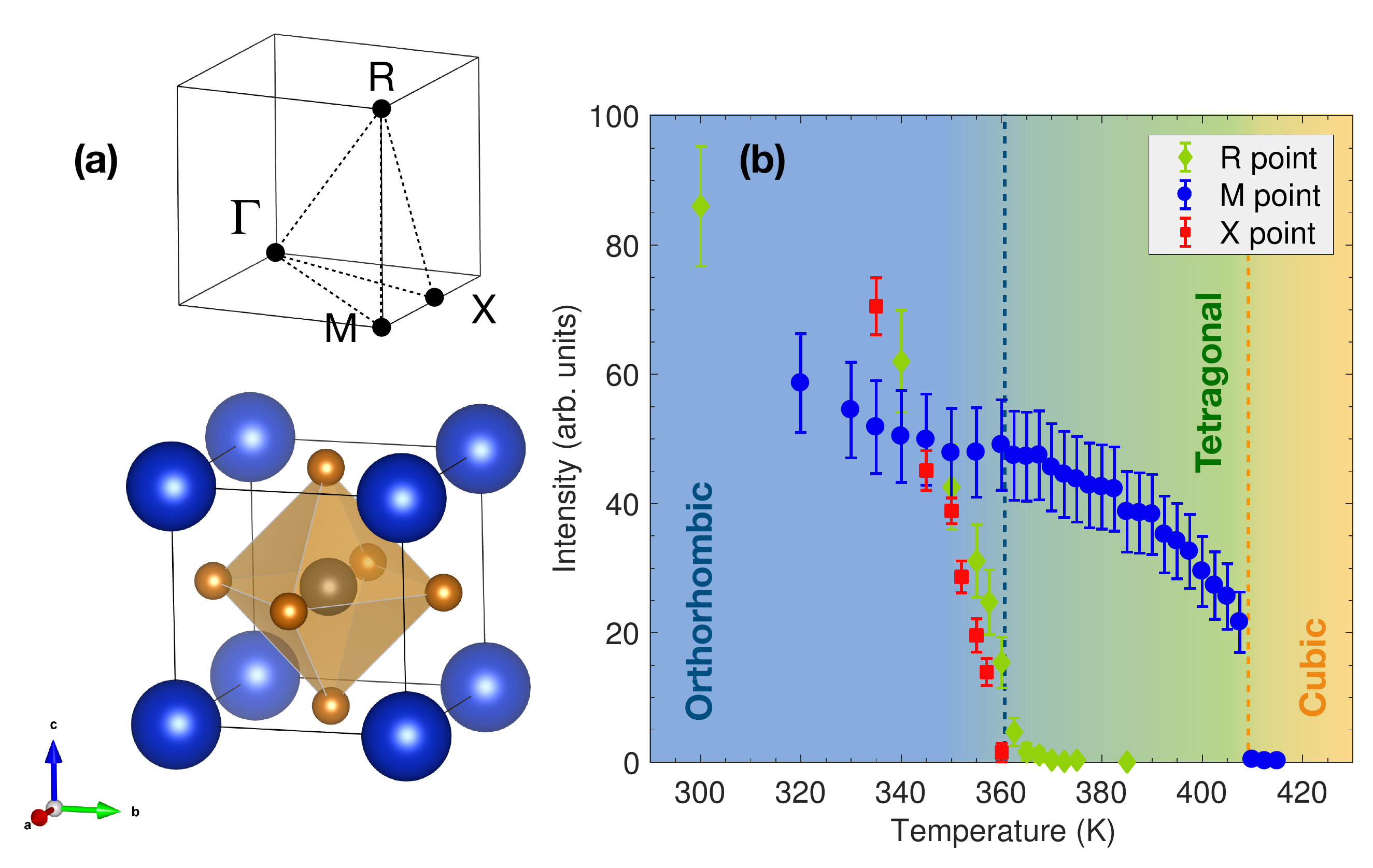}
 \caption{ \label{fig:structure} $(a)$ Crystallographic structure of CsPbBr$_3$ in the cubic phase. The Pb site is represented with the gray sphere, the Br ions are in light brown and the Cs ions are in blue. The figure on the top shows the high symmetry points in the first Brillouin zone of a primitive cubic lattice. $X$, $M$ and $R$ represent $\textbf{Q}=(\frac{1}{2},0, 0), (\frac{1}{2},\frac{1}{2}, 0)$, and $(\frac{1}{2},\frac{1}{2}, \frac{1}{2})$ respectively. $(b)$ Temperature dependence of the superlattice intensities at the three zone boundaries $X$ (2,$\frac{1}{2}$,0), $M$ ($\frac{3}{2}$,$\frac{1}{2}$,0) and $R$ ($\frac{1}{2}$,$\frac{1}{2}$,$\frac{3}{2}$). The colored dashed lines indicate the two transition temperatures.}
\end{figure} 

\section{Experimental details}
\textit{Sample Preparation:} Single crystals of CsPbBr$_3$ were prepared using the Bridgman method. First, a powder of CsPbBr$_3$ was synthesized by solid state reaction following the procedure described in \cite{Stoumpos2013}. A bright orange ingot resulted from the reaction and was ground to obtain a homogeneous powder of CsPbBr$_3$. The powder was then pelleted and sealed in an evacuated quartz ampoule which was placed into a 3-zone horizontal tube furnace. The hottest zone was set to 600$^{\circ}$C and a temperature gradient of 10$^{\circ}$C/cm was applied. Orange transparent crystals were obtained and checked with neutron Laue diffraction using OrientExpress (ILL, Grenoble).

\textit{Neutron powder diffraction measurements:} Powder neutron diffraction measurements were performed on the BT1 diffractometer (NIST, Gaithersburg) using Cu(311) ($\lambda$~=~1.5399 \AA) and Ge(311) ($\lambda$~=~2.079 \AA) monochromators for the orthorhombic and tetragonal phases, respectively.  \textsc{Rietan} \cite{Izumi2007} was used to perform Rietveld refinement of powder neutron diffraction data. Variable temperature synchrotron X-ray powder diffraction data were collected at the Swiss-Norwegian beamline BM01 (SNBL) at the ESRF (France) with an incident wavelength of $\lambda$~=~0.956910 \AA. The Rietveld refinements of this synchrotron X-ray powder diffraction data were performed using the \textsc{Fullprof} suite \cite{Fullprof}.

\textit{Neutron inelastic measurements:} Neutron inelastic spectroscopy was performed on the thermal triple-axis spectrometer IN22 (ILL, Grenoble) with constant $k_f$ = 2.662 \AA$^{-1}$, using a PG filter between the sample and the analyzer to remove higher order contamination. Further measurements as a function of temperature were performed on the thermal triple-axis instrument BT4 (NIST, Gaithersburg) with constant $k_f$~=~2.662~\AA$^{-1}$ using PG filters between the monochromator and the sample, and between the sample and the analyzer. Higher resolution measurements were also carried out on the cold triple-axis spectrometer SPINS (NIST, Gaithersburg) with $k_f$~=~1.55 \AA$^{-1}$ using a Be filter between the sample and the analyzer. 

All phonon measurements were performed in the (H~K~0) and (H~H~L) scattering planes. As super-lattice reflections are expected at the $X$, $M$ and $R$ zone boundaries, measurements were performed towards specific directions where the structure factor was non-zero, as described in \cite{Fujii1974}.  Throughout the paper we use the zone boundary notation taken with respect to the cubic unit cell shown in Fig. \ref{fig:structure} $(a)$.  Diffraction measurements investigating the three zone boundaries were done both in the (H~K~0) and (H~H~L) planes. With the 0.5 g sample aligned in the (H~K~0) plane, the phonon dispersions of the TA$_1$ and TA$_2$ modes could be extracted and another sample aligned in the (H~H~L) plane was used to measure the phonons toward the  $\textbf{Q}_R=(\frac{1}{2},\frac{1}{2}, \frac{3}{2})$ zone boundary. The TA$_1$ mode corresponds to the acoustic phonons propagating along the [1~0~0] direction with a polarization along [0~1~0] or [0~0~1]. Using the equations of motion outlined in Ref. \onlinecite{Dove1993}, the velocity of this phonon can be related to the $C_{44}$ elastic constant. The TA$_2$ phonon mode propagates along [1~1~0] with a polarization along [1~$\overline{1}$~0], and the $lim_{q\rightarrow 0}$ slope of the dispersion depends on ($C_{11}-C_{12}$)/2. Given that the neutron cross section for phonon scattering scales as $(\vec{Q}\cdot \vec{e})^2$, where $\vec{e}$ is the phonon eigenvector, transverse scans near $\vec{Q}=(2~0~0)Ê\pm (0~q~0)$ allows the measurement of the TA$_1$ phonon while scans near $\vec{Q} = (2~2~0) \pm (-q~q~0)$ give the TA$_2$ mode. 

\section{Phonon momentum dependence - frequency and lifetimes}

\begin{figure}
 \includegraphics[scale=0.4]{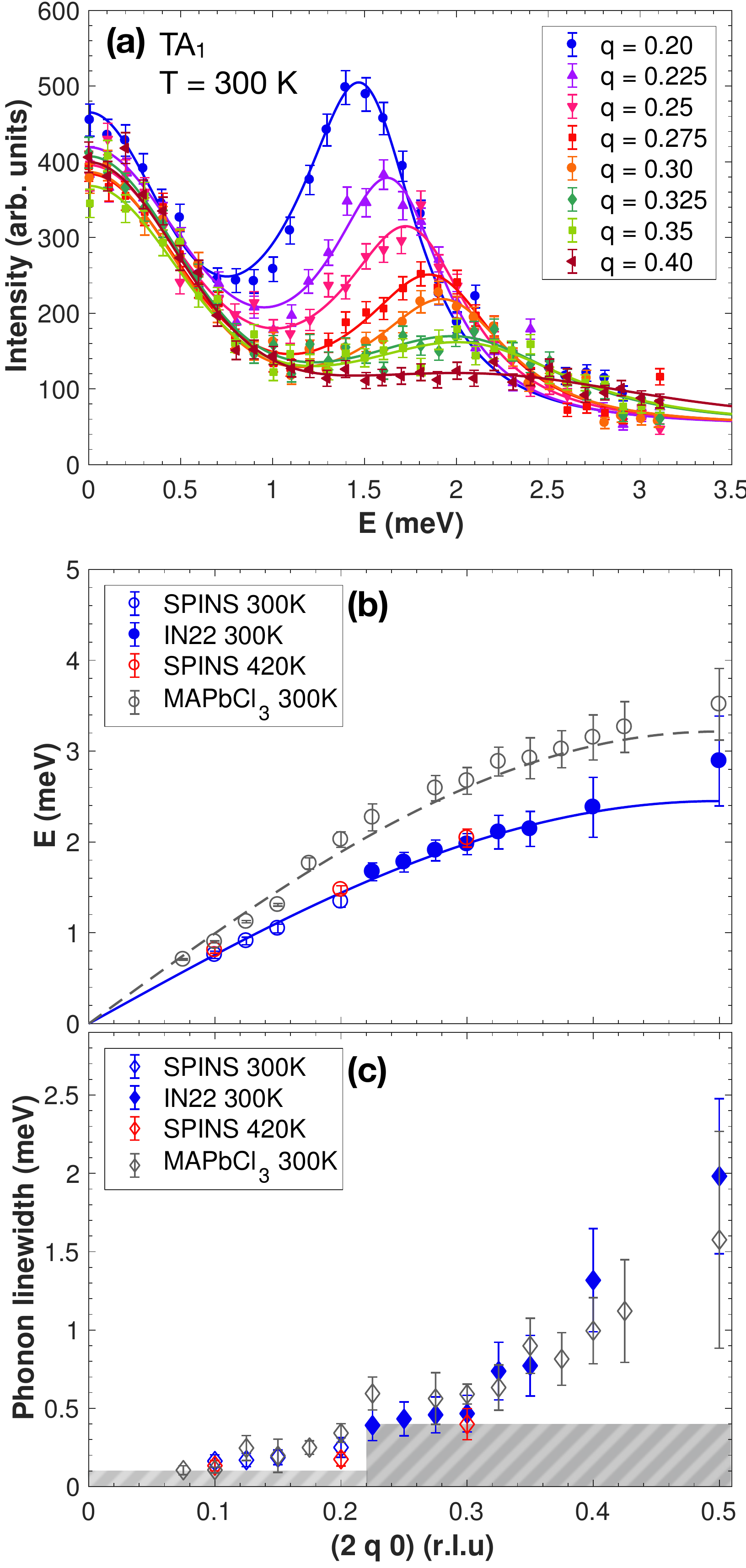}
 \caption{ \label{fig:dispersion_TA1} $(a)$ Constant-Q cuts through the acoustic phonon TA$_1$ for several Q positions from the (2~0~0) Bragg peak towards the  $X$ point. $(b)$ Dispersion curves associated to the TA$_1$ mode towards the $X$ point. The data is compared to the organic-inorganic compound CH$_3$NH$_3$PbCl$_3$ in its cubic phase (gray circles, from \cite{Songvilay2018}). The blue and grey dashed lines are a fit to a sine function performed near the $q \to 0$ limit. $(c)$  Q-dependence of the TA$_1$ phonon linewidths extracted from the constant-Q cuts as described in the text and compared to the phonon linewidth for CH$_3$NH$_3$PbCl$_3$ (grey diamonds, from \cite{Songvilay2018}). The grey dashed areas represent the instrumental resolution of SPINS ($q < $ 0.22) and IN22 ($q >$ 0.22)}
\end{figure} 

\begin{figure}
 \includegraphics[scale=0.4]{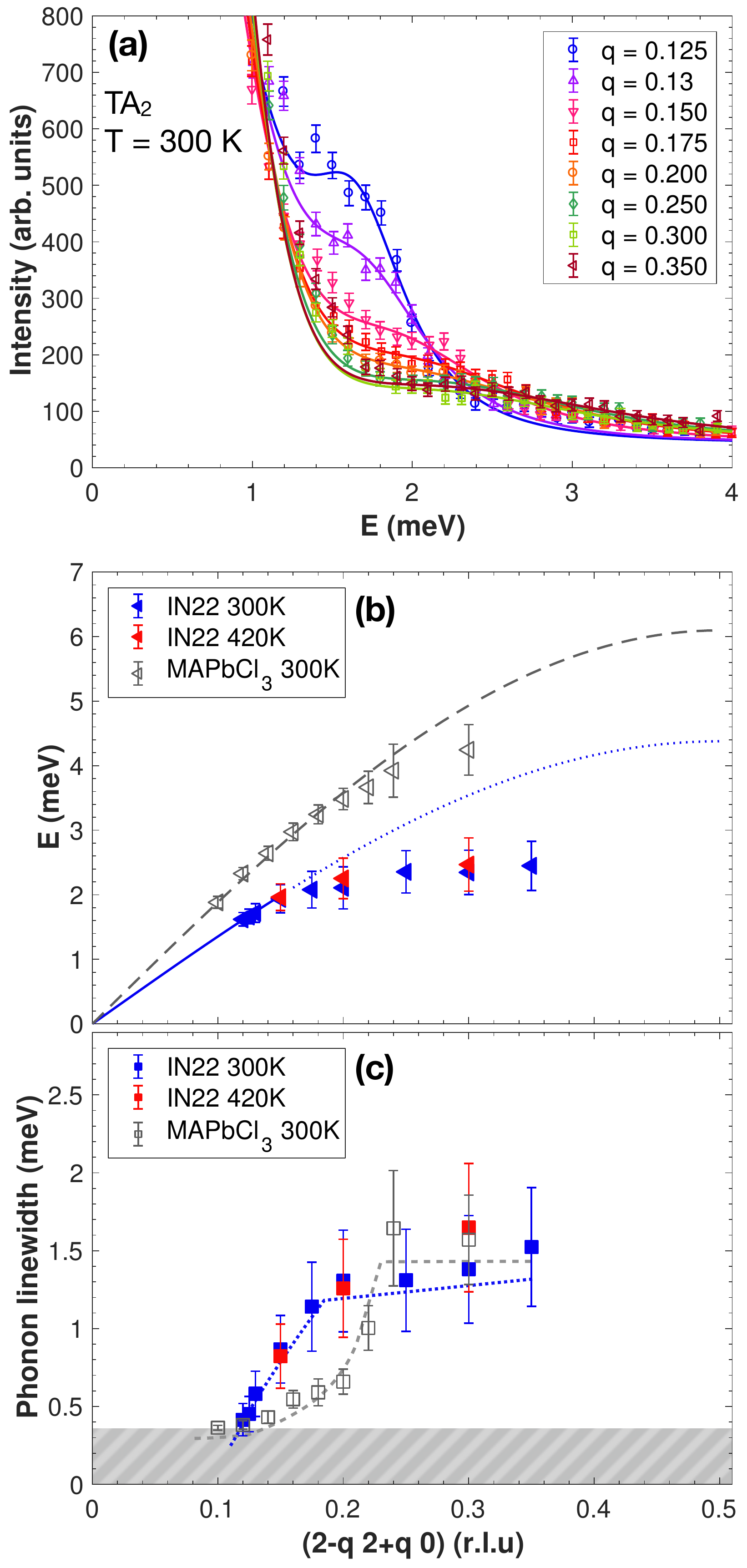}
 \caption{ \label{fig:dispersion_TA2} $(a)$ Constant-Q cuts through the acoustic phonon TA$_2$ for several Q positions from the (2~2~0) Bragg peak towards the  $M$ point. $(b)$ Dispersion curves associated with the TA$_2$ mode towards the $M$ point. The data is compared to the organic-inorganic compound CH$_3$NH$_3$PbCl$_3$ in its cubic phase (grey triangles, from \cite{Songvilay2018}). The blue and grey dashed lines are a fit to a sine function performed near the $q \to 0$ limit. $(c)$  Q-dependence of the TA$_2$ phonon linewidths extracted from the constant-Q cuts as described in the text and compared to the phonon linewidth for CH$_3$NH$_3$PbCl$_3$ (grey squares, from \cite{Songvilay2018}). The blue and grey dashed lines are a guide to the eye. The grey dashed area represents the instrumental resolution of IN22.}
\end{figure} 

Acoustic phonons were mapped out in the (H~K~0) scattering plane and we report in this section anomalies found in the phonon dispersion at particular points of the Brillouin zone. The phonon dispersions were measured in both the orthorhombic and cubic phases at 300~K and 420~K respectively, around the (2~0~0) and (2~2~0) Bragg positions in the \lbrack 2~$q$~0\rbrack\ and  \lbrack 2-$q$~2+$q$~0\rbrack \ directions. These correspond to the directions from the $\Gamma$ zone center point towards the $X$ and $M$ Brillouin zone boundary symmetry points, as shown in Fig. \ref{fig:structure}$(a)$. Dispersions in both directions were measured on the thermal triple-axis spectrometer IN22 (ILL) and were extended to low momentum transfer, to approach the $q \to 0$ limit, on the cold-triple-axis spectrometer SPINS (NIST) in the \lbrack 2~$q$~0\rbrack ~direction. The \lbrack 2-$q$~2+$q$~0\rbrack\ direction is not reachable with cold neutrons due to kinematic constraints of neutron scattering. Figures \ref{fig:dispersion_TA1} $(a)$ and \ref{fig:dispersion_TA2} $(a)$ show constant-Q scans in the \lbrack 2~$q$~0\rbrack~ and \lbrack 2-$q$~2+$q$~0\rbrack~ directions, respectively, at T = 300~K, measured on IN22 with thermal neutrons. Harmonic phonon modes can be observed and show a dependence in momentum transfer. The phonon energy position $\omega_0$ could be extracted as a function of $q$ by fitting the experimental data using a damped harmonic oscillator model and a constant background:
 
 \begin{eqnarray}
 S(\vec{Q},\omega) = ... \nonumber \\
  \left[1 + n(\omega)\right]I_0 \left(\frac{\gamma_0}{\gamma_0^2 + (\omega-\omega_0)^2} -  \frac{\gamma_0}{\gamma_0^2 + (\omega + \omega_0)^2}\right) ,\nonumber
 \end{eqnarray}

\noindent where $\left[1 + n(\omega)\right]$ is the Bose factor, $I_0$ is a constant and $\gamma_0$ is the phonon energy linewidth related to the phonon lifetime via $\gamma_{0} \sim \frac{1}{\tau}$. The above expression of the neutron scattering cross section takes into account both neutron energy gain and loss and obeys detailed balance~\cite{Shirane2004}. The elastic scattering was fitted to a Gaussian centered around E=0 with a width fixed to the instrument resolution extracted from vanadium incoherent scattering.

Fig. \ref{fig:dispersion_TA1} $(b)$ and \ref{fig:dispersion_TA2} $(b)$ show the resulting dispersions from the zone center $\Gamma$ to the $X$ zone boundary (TA$_1$ mode) and $\Gamma$ to $M$ (TA$_2$ mode), respectively. Near the (2~0~0) and (2~2~0) zone centers, the phonons were found to be well defined in energy and momentum. A direct comparison with the phonon dispersion in the organic-inorganic counterpart CH$_3$NH$_3$PbCl$_3$ in each direction (taken from \cite{Songvilay2018}), measured on the IN22 spectrometer and  extracted following the same fitting procedure, is shown in grey.  At low momentum transfer, near the $q \to 0$ limit, the dispersion curves were fitted to a sine function in order to extract the elastic constants $C_{44}$  from the TA$_1$ mode and $(C_{11}-C_{22})/2$  from the TA$_2$ mode. Table \ref{tab:coeff} summarizes the extracted elastic constant values along with a comparison with the chlorine counterpart and the hybrid compound CH$_3$NH$_3$PbCl$_3$. The elastic constants for our Br variant are generally lower than the Cl and organic-inorganic compounds.  Given the larger size of Br, the decrease may be attributed to an increase in lattice constants. It can be noted that no clear change in the dispersion curves occur between 300~K and 420~K on the energy scale ($\sim$ THz) probed with neutron scattering. The temperature dependence of the lattice dynamics will be discussed later. Moreover, the TA$_2$ mode becomes significantly flat, dispersing little in energy towards the $M$ zone boundary.

While acoustic phonons are well defined near the zone center with $\gamma_0 < \omega_{0}$, they become much broader in energy towards the zone boundaries, indicative of a shortened lifetime. The phonon linewidth was extracted as a function of $q$ and shows a shorter lifetime for smaller wavelength excitations, as shown in Figures \ref{fig:dispersion_TA1} $(c)$ and \ref{fig:dispersion_TA2} $(c)$. The effect is the most dramatic for the TA$_2$ phonon mode, towards the $M$ point as phonons could not be observed above $q$ = 0.35. This feature was also reported in CsPbCl$_3$ \cite{Fujii1974}. As shown in grey, the momentum dependence of acoustic linewidth is also very similar to the organic-inorganic hybrid perovskite CH$_3$NH$_3$PbCl$_3$ \cite{Songvilay2018} near the zone boundary.  At the momentum transfers near the zone center, the linewidth enters the resolution of the spectrometer and higher resolution probes are required to investigate the phonon lifetime in this region of momentum.

\begin{table*}[t]
\caption{\label{tab:coeff}Comparison of elastic constants extracted from ultrasound measurements in the cubic phase for CsPbCl$_3$ and neutron inelastic measurements in the cubic phase for CsPbBr$_3$ and MAPbCl$_3$ (MA = CH$_{3}$NH$_{3}$).}
\begin{ruledtabular}
\begin{tabular}{llll}
& CsPbCl$_3$ (from \cite{Hirotsu1978}) & CsPbBr$_3$ (this work) & MAPbCl$_3$ (from \cite{Songvilay2018})  \\
 \hline
 $C_{44}$ (GPa)  & 5.04  & 2.45(3) & 3.00 \\
 ($C_{11}-C_{12}$)/2 (GPa) & 8.95 & 4.45(5) & 10.8 \\
\end{tabular}
\end{ruledtabular}
\end{table*}


\section{Temperature dependence}

\subsection{Static properties}

A temperature dependence study was carried out to characterize the different structural transitions in CsPbBr$_3$. The intensity of the elastic superlattice reflections at the $X$ point (2,$\frac{1}{2}$,0), $M$ point ($\frac{3}{2}$,$\frac{1}{2}$,0) and $R$ point ($\frac{1}{2}$,$\frac{1}{2}$,$\frac{3}{2}$) was measured as a function of temperature on the IN22, BT4 and SPINS spectrometers, respectively. New nuclear Bragg peaks appear at all three zone boundaries, implying a doubling of the unit cell along the crystallographic directions on entering the tetragonal and orthorhombic phases from the high temperature cubic phase. As shown in Fig. \ref{fig:structure}$(b)$, a sudden jump in the intensity occurs at the $M$ point around T = 410~K when the unit cell transforms from cubic to tetragonal.  When entering the low-temperature orthorhombic phase around T~=~360~K, superlattice intensities simultaneously appear at the $X$ and $R$ points. While the discontinuous evolution of the intensity at the $M$ point suggests a first-order transition, the intensity at the $X$ and $R$ points seem to follow a second-order transition, as previously reported in \cite{Hirotsu1974,Hua1991}. 

\begin{figure}
 \includegraphics[scale=0.45]{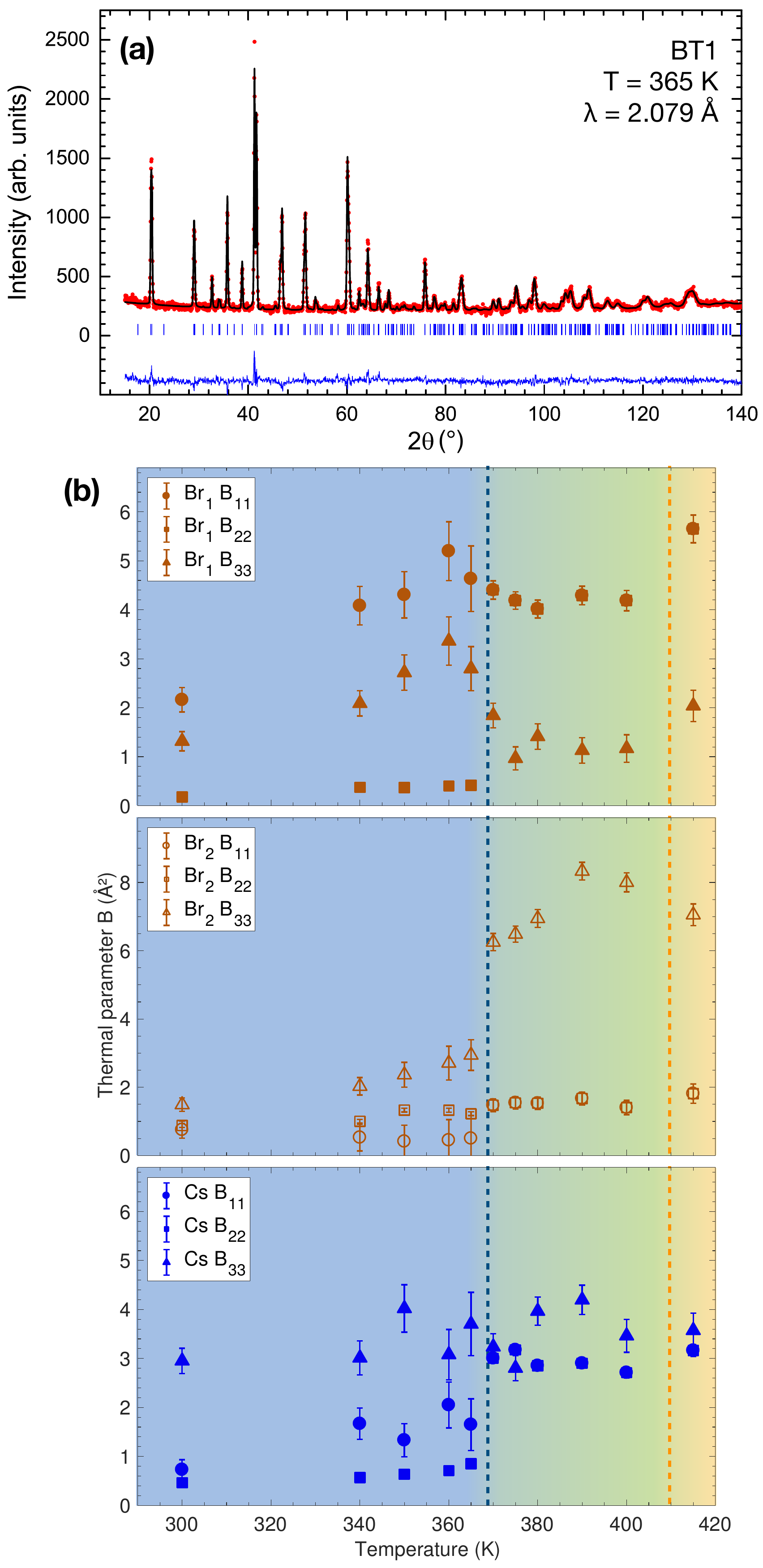}
 \caption{ \label{fig:diffraction} $(a)$ Powder neutron diffraction data performed on BT1 ($\lambda$ = 2.079 \AA) at T = 365 K. The red dots and black line represent the measured and calculated intensities, respectively. The blue stick marks indicate the calculated reflections and the blue line represents the difference between measured and calculated intensities. $(b)$ Thermal parameters for the three spatial components associated with the Br$_1$, Br$_2$ and Cs sites, respectively, as a function of temperature. The colored areas and dashed lines indicate the tetragonal and orthorhombic transitions.}
\end{figure}

\begin{figure}
 \includegraphics[scale=0.4]{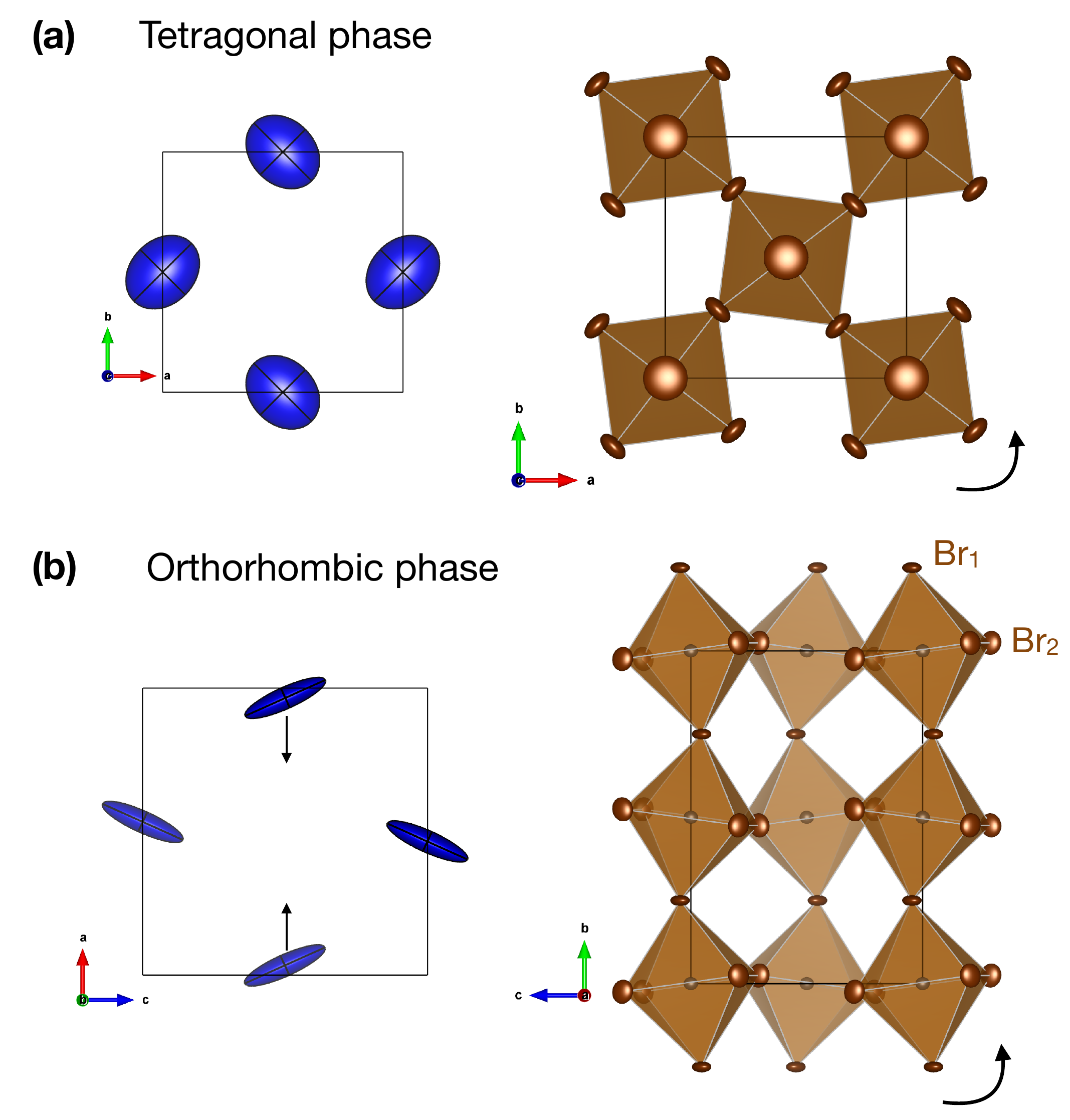}
 \caption{ \label{fig:tilt} $(a)$-$(b)$ Thermal displacements of Cs ions (blue ellipsoids) and PbBr$_6$ octahedra tiltings (right panels) in the tetragonal and orthorhombic phases, respectively. The arrows indicate atomic displacements and octahedral rotations.}
\end{figure}

Previous work on CsPbCl$_{3}$ analyzing the structure found irregular behavior on the Cs and Cl sites which could be modelled as either an anomalously large thermal parameter~\cite{Harada1976} or site disorder~\cite{Moller1959}.  Following this, we have carried out an analysis of the temperature dependence of the thermal parameters for CsPbBr$_{3}$ using powder neutron diffraction (Fig. \ref{fig:diffraction}) and synchrotron X-ray diffraction. The refined values as a function of temperature, for the components of the thermal parameters along the three directions in space ($B_{11}$, $B_{22}$ and $B_{33}$) for Br and Cs are displayed in Figure \ref{fig:diffraction} $(b)$ and are illustrated in real space in Figure \ref{fig:tilt}. The blue ellipsoids in the left panels of Fig. \ref{fig:tilt} represent thermal displacements of the Cs ions in the tetragonal and orthorhombic phases and the right panels show the octahedral tiltings associated with both structural transitions. It should be noted that the powder diffraction data showed a slightly different temperature for the transition from the tetragonal to orthorhombic phase (found around 365~K), compared to the single crystal sample (360 K). Like in CsPbCl$_{3}$, the thermal parameters values are anomalously large for the Br and Cs sites across all temperatures. 

In the cubic phase, Figure \ref{fig:diffraction} $(b)$ (bottom panel) shows that the Cs motions are quasi-isotropic while the Br motions (top and middle) are large and anisotropic in comparison, as also previously reported in \cite{Harada1976} for CsPbCl$_3$. When going through the tetragonal and orthorhombic transition however, the Cs site also becomes anisotropic as indicated by the elongated shapes of the thermal ellipsoids in Fig. \ref{fig:tilt}. Moreover, as illustrated in Figure \ref{fig:tilt} $(b)$, the results also show an off-center rearrangement of the A-site cation in the low-temperature phase, indicated by the arrows. 
The shape and orientation of the ellipsoids representing the Cs thermal displacements follows the PbBr$_6$ octahedra tilting in both tetragonal and orthorhombic phases, which suggests a strong coupling between the A-site cation and the halogen anions, as discussed later.

\subsection{Lattice dynamics}

The temperature evolution of the lattice dynamics was also investigated. As mentioned and shown in Fig. \ref{fig:dispersion_TA1} and \ref{fig:dispersion_TA2}, no clear change in the phonon dispersion curves for CsPbBr$_{3}$ between the high-temperature cubic phase and the low-temperature orthorhombic phase, was observed near the zone center. This contrasts with the case of CH$_{3}$NH$_{3}$PbCl$_{3}$ where the elastic constants, and hence the acoustic phonon dispersion, shows a discontinuity at the low temperature structural transitions.~\cite{Songvilay2018}  This may be indicative of an underlying ferroelastic character as suggested by ultrasound measurements.~\cite{Harwell2018} 

\begin{figure}
 \includegraphics[scale=0.4]{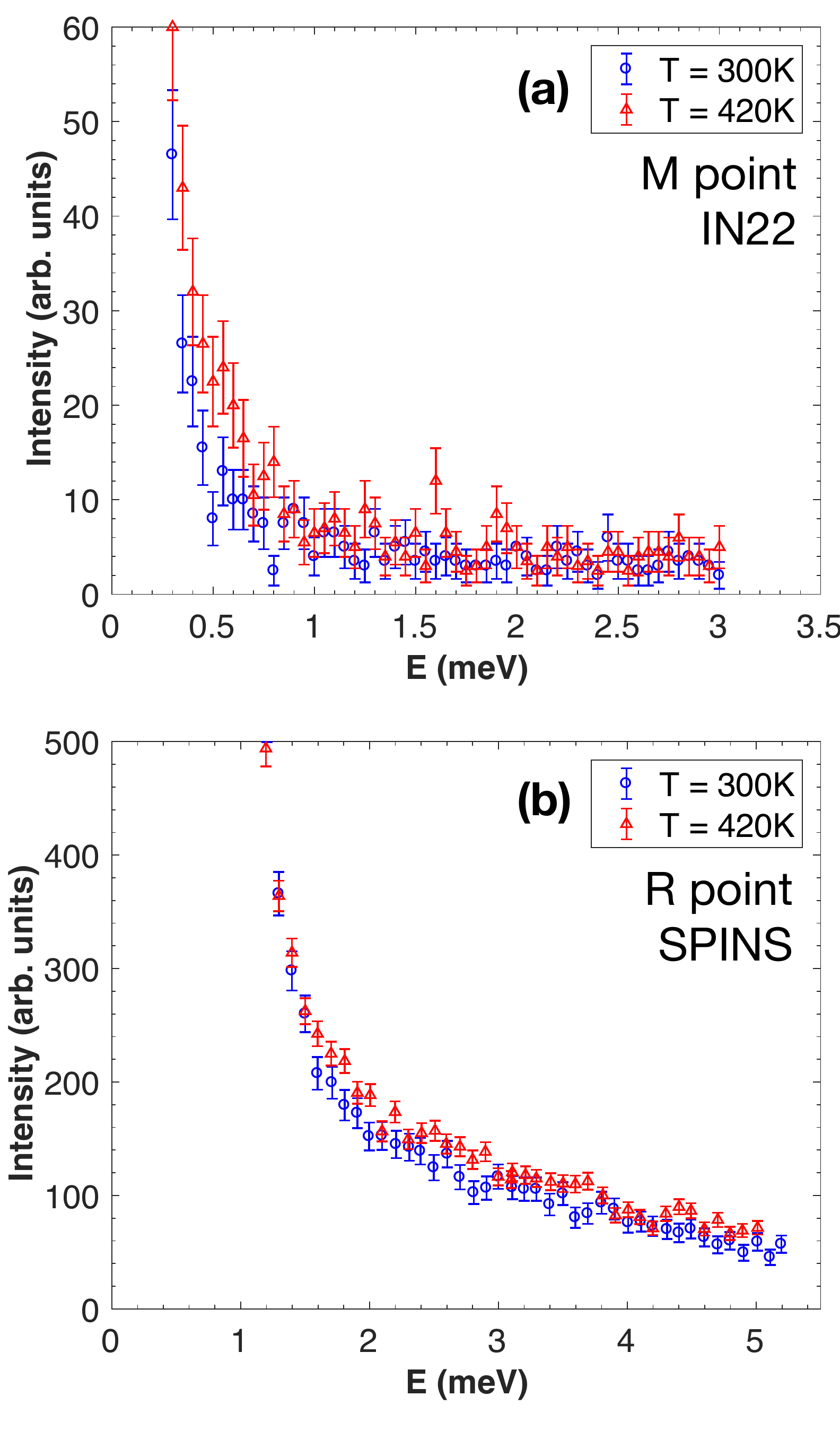}
 \caption{ \label{fig:temp_RM} $(a)$ Constant-Q cuts through the TA$_1$ mode at the $M$ point ($\frac{3}{2}$,$\frac{1}{2}$,0) measured on IN22 at 300~K and 420~K.   $(b)$ Constant-Q cuts through the TA$_1$ mode at the $R$ point ($\frac{1}{2}$,$\frac{1}{2}$,$\frac{3}{2}$) measured on SPINS at 300~K and 420~K.}
\end{figure}

\begin{figure}
 \includegraphics[scale=0.4]{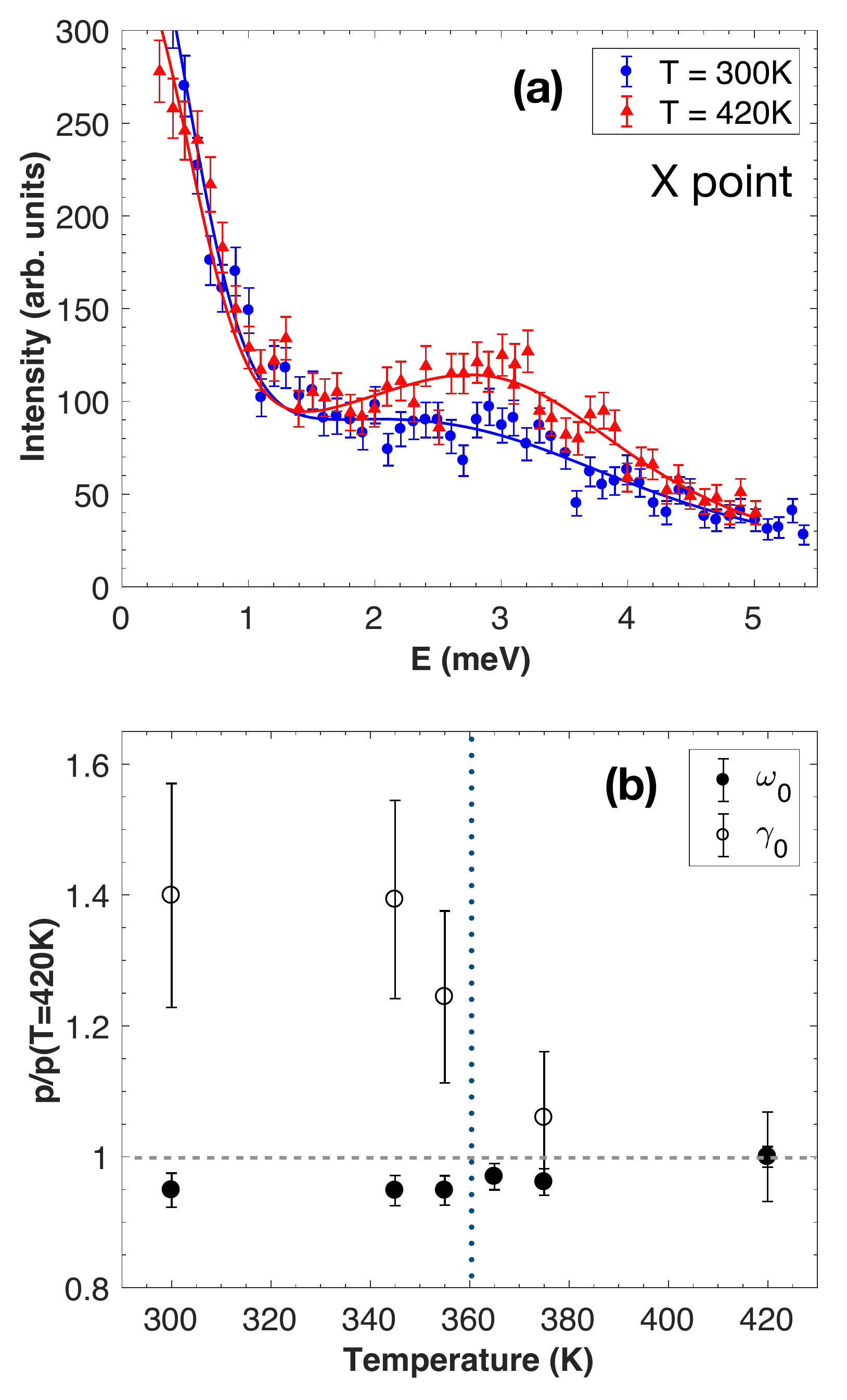}
 \caption{ \label{fig:temp_X} $(a)$ Constant-Q cuts through the TA$_1$ mode at the $X$ point measured on IN22 at 300~K (orthorhombic phase) and 420~K (cubic phase). $(b)$ Temperature dependence of the energy position (black circles) and the energy linewidth (white circles) of the TA$_1$ phonon at the $X$ (2,$\frac{1}{2}$,0) zone boundary. The vertical dashed line indicates the orthorhombic transition.}
\end{figure}

We now discuss the temperature dependence at the $X$ (2,$\frac{1}{2}$,0) and $M$ ($\frac{3}{2}$,$\frac{1}{2}$,0) zone boundaries, through the two structural transitions. Figures \ref{fig:temp_RM} $(a)$-$(b)$ and \ref{fig:temp_X} $(a)$ show constant-Q scans at the $M$, $R$, and $X$ points respectively in the cubic (T = 420~K) and orthorhombic (T = 300~K) phases. While the acoustic phonons remains unobservable at the $M$ and $R$ points in both phases, the TA$_1$ phonon shows a broadening at the $X$ point when cooling into the orthorhombic phase.  As also reported in CsPbCl$_3$ \cite{Fujii1974}, the presence of highly damped phonons around the $M$ and $R$ points does not allow for a direct measurements of the soft modes at these zone boundaries throughout the transitions. Fig. \ref{fig:temp_X} $(b)$ displays the relative change in linewidth $\gamma_0$ and energy position $\omega_0$ of the TA$_1$ mode, starting from the high temperature phase (set as the reference) and cooling down towards 300~K. With decreasing temperature, the TA$_1$ mode seems to soften very slightly while there is a clear increase in the phonon linewidth when entering the orthorhombic phase, indicating a shortening of the phonon lifetime. 

\begin{figure}
 \includegraphics[scale=0.36]{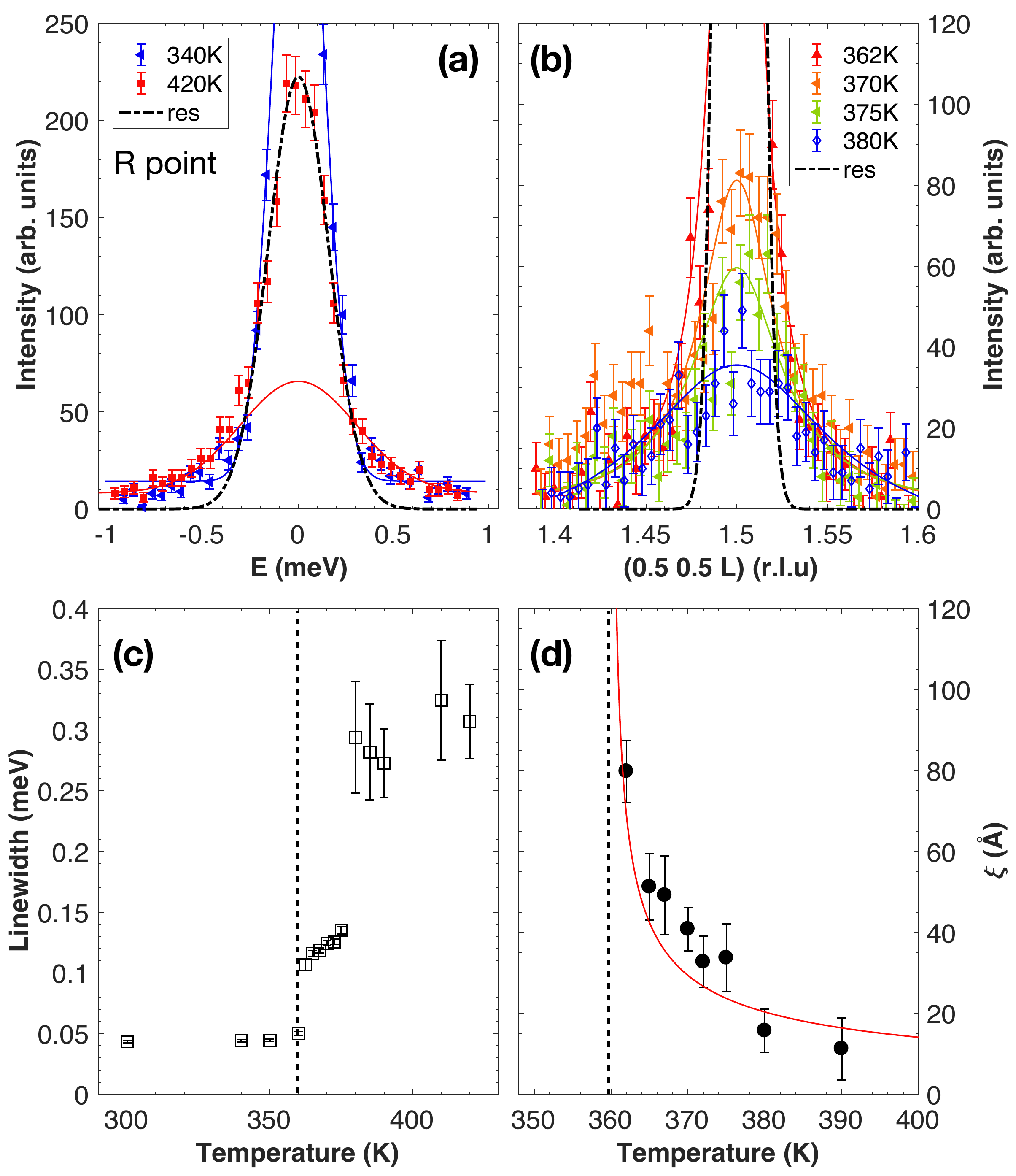}
 \caption{ \label{fig:crit_scat_R} $(a)$ Energy scans measured on SPINS at the $R$ point ($\frac{1}{2}$,$\frac{1}{2}$,$\frac{3}{2}$) in the cubic and orthorhombic phases. The black dashed line indicates the instrumental energy resolution, extracted from the incoherent signal of vanadium. $(b)$ Elastic scans measured on SPINS at the $R$ point ($\frac{1}{2}$,$\frac{1}{2}$,$\frac{3}{2}$) at several temperatures. The black dashed line shows the Bragg instrumental resolution in momentum. $(c)$ Temperature dependence of the energy linewidth extracted from the quasi-elastic scans at the $R$ point. $(d)$ Temperature dependence of the spatial correlation (inverse of the momentum linewidth) extracted from the elastic scans at the $R$ point. The vertical dashed line indicates the temperature transition to the orthorhombic phase.}
\end{figure} 

\begin{figure}
 \includegraphics[scale=0.35]{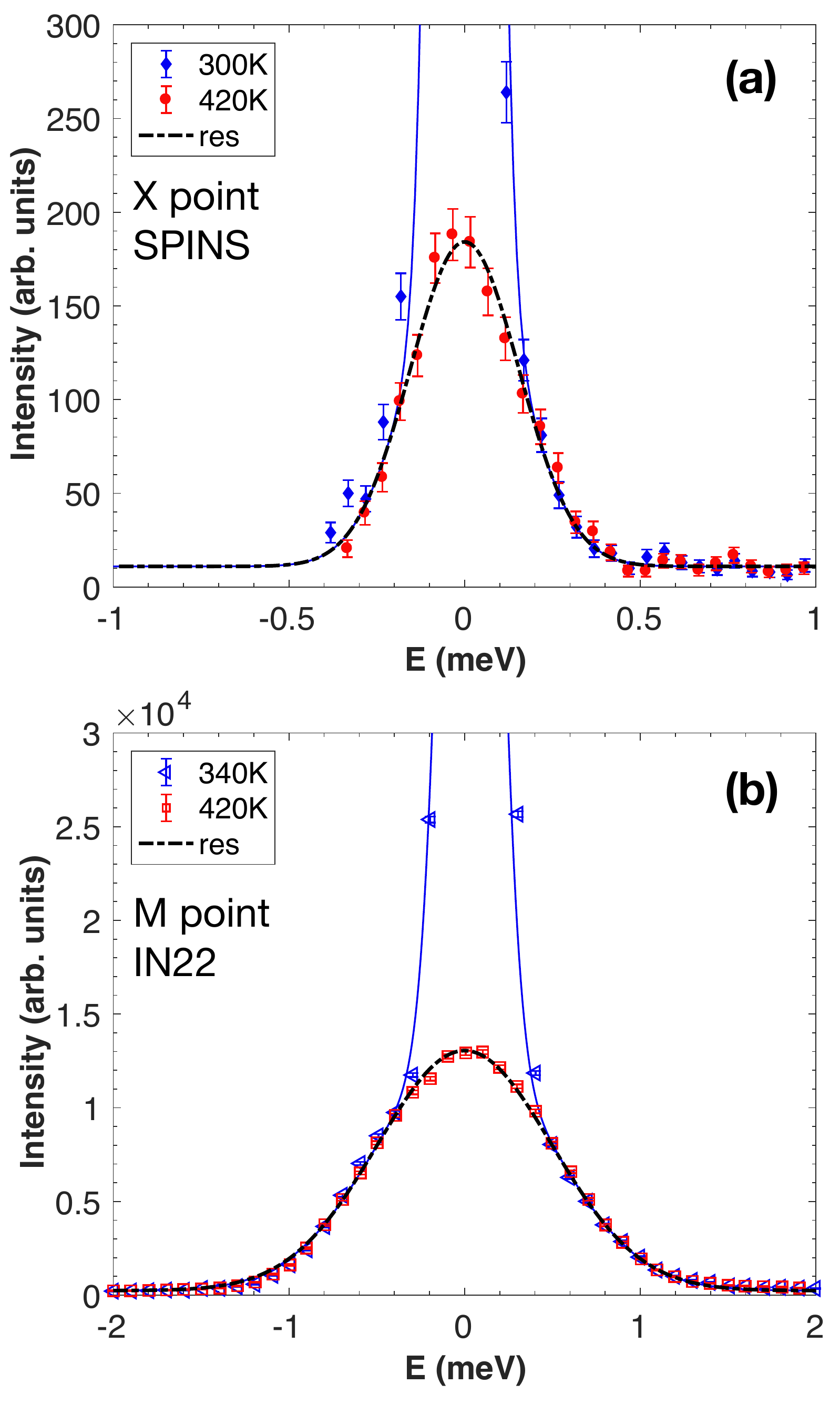}
 \caption{ \label{fig:crit_scat_XM} $(a)$-$(b)$ Energy scans measured on SPINS and IN22 at the $X$ and $M$ points, respectively, in the cubic and orthorhombic phase.}
\end{figure}

We now discuss the relaxational scattering around E~=~0 meV for the three zone boundaries $X$ (2,$\frac{1}{2}$,0), $M$ ($\frac{3}{2}$,$\frac{1}{2}$,0) and $R$ ($\frac{1}{2}$,$\frac{1}{2}$,$\frac{3}{2}$). Energy scans were carried out as a function of temperature, using the cold triple-axis spectrometer SPINS for the measurements at the $R$ and $X$ points (Fig. \ref{fig:crit_scat_R} $(a)$ and \ref{fig:crit_scat_XM} $(a)$ respectively) and the thermal triple-axis spectrometer IN22 for the measurements at the $M$ point (Fig. \ref{fig:crit_scat_XM} $(b)$). At the $X$ and $M$ zone boundaries, the data shows a rapid increase in the intensity when decreasing the temperature, while the width in energy remains resolution limited. It should be noted that the low temperature data corresponds to the appearance of new Bragg peaks, with resolution tied to the instrumental Bragg resolution which is much narrower than the resolution extracted from the incoherent signal of vanadium as shown with the black dashed line. In contrast, at the $R$ point, critical dynamic fluctuations beyond the instrumental resolution could be observed above the tetragonal-to-orthorhombic phase transition. 

In order to parameterize the observed neutron scattering cross section, the experimental data was fitted to the sum of a Gaussian with width fixed to the experimental resolution derived from the incoherent scattering from vanadium (shown with black dashed lines) and a second narrower Gaussian function. Fig. \ref{fig:crit_scat_R} $(c)$ shows the temperature evolution of the extracted energy linewidth. In particular, the linewidth shows a first jump when decreasing temperature from the cubic to the tetragonal phase. Then, when the system enters the orthorhombic phase, the linewidth further decreases and reaches a minimum value corresponding to the Bragg resolution while the intensity increases. The presence of critical fluctuations in energy is concomitant with a narrowing in momentum as observed with elastic transverse scans shown in Fig. \ref{fig:crit_scat_R} $(b)$ when decreasing temperature. The elastic scattering, which is broad in momentum above the cubic phase, progressively sharpens along with an increase in intensity, indicating a critical divergence of the dynamic correlation length. The temperature dependence of the correlation length $\xi$ was extracted by fitting the elastic data with a Lorentzian of the form:

\begin{eqnarray}
 I(q) = \frac{I_0\gamma}{(\gamma^2 + q^2)} ,
 \end{eqnarray}
where $I_0$ is a constant, $\gamma = 1/\xi$ is the inverse of the real-space correlation length and $q$ is the length of the wave-vector measured from the ($\frac{1}{2}$,$\frac{1}{2}$,$\frac{3}{2}$) position.
The spatial correlation length, plotted as a function of temperature in Fig. \ref{fig:crit_scat_R} $(d)$, shows a divergence when going towards the orthorhombic transition at T = 360 K, as indicated by the dashed line. A power-law fit of the correlation length, following a mean-field analysis, as a function of temperature using a form $\xi(T) = \xi_0\left(\frac{T-Tc}{Tc}\right)^{\nu}$ was also performed (shown with the red line) and gave an exponent of $\nu$ = -0.53 $\pm$ 0.06, close to the exact value $-\frac{1}{2}$ predicted from mean field theory~\cite{Collins:book}.

\section{Discussion}

\subsection{Energy broadening of the TA$_2$ mode}

We first address the significant energy broadening of the phonons towards the $M$ zone boundary of the TA$_2$ phonon mode. Temporally broad acoustic phonons have also been reported for the hybrid organic-inorganic halide perovskites CH$_3$NH$_3$PbCl$_3$\cite{Songvilay2018} and CH$_3$NH$_3$PbI$_3$\cite{GoldParker2018} and attributed to the molecular dynamics~\cite{Brown2017,Swainson2015} strongly affecting the acoustic phonon modes through hydrogen bonding. It was further suggested~\cite{GoldParker2018}, with lattice dynamics calculations, that these anharmonic effects are due to low-energy optical phonon modes arising from the organic cation.  The resulting short mean free path of the phonon resulting from this dampening was suggested to be the origin of the ultra-low thermal conductivity in this compound.

Figures \ref{fig:dispersion_TA1} and \ref{fig:dispersion_TA2} illustrate that this same phonon damping is present in CsPbBr$_{3}$, even without the presence of an organic molecule.  However, in the case of CsPbBr$_3$, the powder diffraction data presented in Fig. \ref{fig:diffraction} shows large anisotropic Cs and Br motions in both the tetragonal and orthorhombic phases. These anisotropic Cs displacements respond in concert with octahedra tilting, suggestive of a strong coupling between the $A$ cation and the halogen anions. The large anisotropic displacements on the $A$-site may, in turn, damp tilt modes of the PbBr$_{6}$ octahedra manifesting as energy broadening near the $M$ zone boundary.  As discussed below, this is consistent with the structure factors of neutron scattering,~\cite{Fujii1974, Lynn1978} and is supported by calculations~\cite{Bechtel2018_1,Yang2017,Marronnier2017} on a series of fully inorganic lead halide perovskites, which have reported anharmonic fluctuations, including octahedral tilting. In particular, Refs. \cite{Bechtel2018_1,Yaffe2017} stress the importance of the coupling between the Cs$^+$ displacements and the tilt distortions in order to explain the dynamical instabilities in the high and intermediate temperature phases. 

While the energy widths of the acoustic phonons are similar in CsPbBr$_3$ and CH$_3$NH$_3$PbCl$_3$, there are some differences between the organic and inorganic perovskite variants. Contrary to the organic compound, the TA$_1$ and TA$_2$ phonon linewidths in CsPbBr$_3$, near the zone center,  do not show any measurable change in temperature (on the THz scale) while the organic variant shows a broadening of the long wavelength acoustic phonons in the intermediate tetragonal phase. This additional temperature dependent broadening may be the result of the molecular motion and this indeed has been suggested by Raman scattering (at $Q\rightarrow0$ probe) with the observation of an ``amplitudon" mode in the organic variant.~\cite{Guo2017}  Therefore, the effects of the organic molecule may be dominant for long wavelength dynamics sampled through phonons near the zone center rather than the zone boundary.  

\subsection{Temperature dependent dynamics and statics near the zone boundaries}

The structural transitions in the lead halide perovskite compounds CsPb(Cl,Br)$_3$ were previously studied by neutron scattering \cite{Fujii1974,Hirotsu1974,Hua1991} and confirmed by our diffraction results in Fig. \ref{fig:structure}. CsPbCl$_3$ exhibits three transitions while the Br counterpart undergoes only two with a cubic to tetragonal transition at $\sim$ 400~K and an orthorhombic unit cell forming below 360~K. Group theory analysis~\cite{Hirotsu1974} predicts that in CsPbBr$_3$, the higher temperature transition is caused by the condensation of the $M_3$ mode at the $M$ point $(\frac{1}{2},\frac{1}{2}, 0)$. The second transition is associated with the $R$ point in the cubic Brillouin zone which transforms into the $Z$ point of the tetragonal Brillouin zone. We note that both $R$ and $X$ points of the cubic lattice are equivalent in the tetragonal phase and hence show simultaneous superlattice reflections.  The $R_{25}$ mode associated to the $R$ point splits into a $Z_1$ mode and a doubly degenerate $Z_9$ mode. It is predicted that the transition is driven by a softening of the $Z_9$ modes which are based on the motions of both the Cs and Br ions. 

As discussed in Ref. \cite{Hua1991}, and in relation to neutron scattering structure factors in \cite{Lynn1978,Gesi1972}, the $M$ point involves the tilting of the PbCl$_6$ octahedra around the \lbrack0~0~1\rbrack\ cubic axis (also shown in the right panel of Fig. \ref{fig:tilt} $(a)$), while the $R$ point (measured in the (H H L) scattering plane) involves displacive distortions of the octahedra accompanied with the displacement of Cs ions. The $A$-site Cs motions may explain the critical scattering (on the THz frequency scale) observed at the $\textbf{Q}_R=(\frac{1}{2},\frac{1}{2}, \frac{3}{2})$ point and presented in Fig. \ref{fig:crit_scat_R} $(a)$-$(d)$, yet simultaneously not measurable at the $\textbf{Q}_X=(2,\frac{1}{2}, 0)$ point.  While both $\textbf{Q}_R$ and $\textbf{Q}_X$ are the same zone boundary (termed $Z$) in the tetragonal phase, they are different momentum points with $\textbf{Q}_X$ having an identical zero structure factor for the Cs site while the Cs $A$-site motion gives a nonzero neutron structure factor at $\textbf{Q}_R$.  This is consistent with the group theory analysis outlined above and the powder diffraction data shown in Fig. \ref{fig:diffraction}.  

Our temperature dependent results are consistent with critical fluctuations of the Cs $A$-site and Fig. \ref{fig:crit_scat_R} $(c)$ illustrates two-steps suggestive of Cs dynamics in both the high temperature cubic phase and the intermediate tetragonal phase.  This is supported by theoretical work~\cite{Bechtel2018_1} highlighting the importance of displacive $A$-site motion and its coupling to octahedra tilt modes.  Concurrently, we note that dynamic critical fluctuations are not observable in CH$_3$NH$_3$PbCl$_3$~\cite{Songvilay2018} at the $M$ and $X$ zone boundaries, but show a sudden increase of resolution limited Bragg peaks.  This may indicate that the transitions in the inorganic variants is dominated by a coupling to relaxational dynamics rather than displacive~\cite{Shapiro1972,Shirane1969,Shirane1993,Halperin1976}.  This later point requires further study with higher resolution techniques.

The comparison between the organic-inorganic perovskites and the fully inorganic variants indicate a strong coupling between the $A$ site and the tilt distortions, even in the absence of hydrogen bonding.  It also appears that the influence of the $A$ site, either disordered or strongly fluctuating, are ubiquitous to the lead halide perovskites.  This in turn results in highly damped acoustic phonon fluctuations and may explain the apparently universal low thermal conductivity found both in fully inorganic \cite{Lee2017} and organic-inorganic lead halide perovskites~\cite{GoldParker2018}.

\section{Conclusion}

We report short acoustic phonon lifetimes in the fully inorganic CsPbBr$_{3}$. The phonon lifetimes are similar to organic variants, indicating that the presence of an organic molecule on the $A$-site and the resulting hydrogen bonding are not crucial for large damping to appear in a large part of the Brillouin zone.

\begin{acknowledgments}
We acknowledge funding from the EPSRC, STFC and Carnegie Trust for the universities of Scotland. The work was also supported by the Natural Sciences and Engineering Research Council of Canada (NSERC, Grant No. 203773) and the U. S. Office of Naval Research (Grant No. N00014-16-1-3106). We acknowledge the support of the National Institute of Standards and Technology, U.S. Department of Commerce, in providing the neutron research facilities used in this work.  
\end{acknowledgments}

\bibliography{CsPbBr_final}

\end{document}